\documentclass[prl,twocolumn,showpacs,superscriptaddress]{revtex4}

\usepackage{graphicx}
\usepackage{epsfig}
\usepackage{verbatim}
\usepackage{amssymb}
\usepackage{amsfonts,amsmath}
\usepackage{bm}
\usepackage[T1]{fontenc}
\usepackage{eufrak}
\usepackage[colorlinks]{hyperref}
\usepackage{fancyhdr}

\hypersetup{colorlinks=true,
	linkcolor=red,
	anchorcolor=green,
	citecolor=blue,
	urlcolor=green
}

\providecommand{\be}{\begin{equation}}
\providecommand{\ee}{\end{equation}}
\providecommand{\bea}{\begin{eqnarray}}
\providecommand{\eea}{\end{eqnarray}}
\providecommand{\beas}{\begin{eqnarray*}}
\providecommand{\eeas}{\end{eqnarray*}}

\providecommand{\beni}{\begin{equation*}}
\providecommand{\eeni}{\end{equation*}}

\providecommand{\bw}{\begin{widetext}}
\providecommand{\ew}{\end{widetext}}

\newlength{\bilderlength}

\newlength{\figsize}
\begin{document}

\title{Extreme Value Statistics Distributions in Spin Glasses}

\author{Michele Castellana}
\email{michele.castellana@lptms.u-psud.fr}
\affiliation{LPTMS, CNRS and Universit\'{e} Paris-Sud, UMR8626, B\^{a}t. 100, 91405 Orsay, France} 
\affiliation{Dipartimento di Fisica, Universit\`a di Roma `La Sapienza' , 00185 Rome, Italy}
\author{Aur\'elien Decelle}
\affiliation{LPTMS, CNRS and Universit\'{e} Paris-Sud, UMR8626, B\^{a}t. 100, 91405 Orsay, France} 
\email{aurelien.decelle@lptms.u-psud.fr}

\author{Elia Zarinelli}
\affiliation{LPTMS, CNRS and Universit\'{e} Paris-Sud, UMR8626, B\^{a}t. 100, 91405 Orsay, France} 
\email{elia.zarinelli@lptms.u-psud.fr}


\begin{abstract} 
We study the probability distribution of the pseudo-critical temperature  in a mean-field and in a short-range spin-glass model: the Sherrington-Kirkpatrick (SK) and the Edwards-Anderson (EA) model. In both cases, we put in evidence the underlying connection between the fluctuations of the pseudo-critical point and and the Extreme Value Statistics of random variables. For the SK model, both with Gaussian and binary couplings, the distribution of the pseudo-critical temperature is found to be the Tracy-Widom distribution. For the EA model, the distribution is found to be the Gumbel distribution. Being the EA model  representative of uniaxial magnetic materials with quenched disorder like  $\text{Fe}_{0.5}\text{Mn}_{0.5}\text{TiO}_3$ or $\text{Eu}_{0.5}\text{Ba}_{0.5}\text{MnO}_3$, its pseudo-critical point distribution should be \emph{a priori} experimentally  accessible. 
\end{abstract}

\maketitle

Disordered uniaxial magnetic materials having a glassy behavior like $\text{Fe}_{0.5}\text{Mn}_{0.5}\text{TiO}_3$ \cite{gunnarson1991static} and $\text{Eu}_{0.5}\text{Ba}_{0.5}\text{MnO}_3$ \cite{nair2007critical} 
have interested physicists for decades. Since the first pioneering work of Edwards and Anderson (EA) \cite{edwards1975theory}, these systems have been studied by means of spin-glass models with quenched disorder, which were later considered in their mean-field version by Sherrington and Kirkpatrick (SK) \cite{sherrington1975solvable}. In the thermodynamic limit, Parisi's solution for the SK model \cite{MPV} predicts a phase transition at a finite critical temperature separating a high-temperature paramagnetic phase from a low-temperature glassy phase. Differently, for the EA model there is no analytical solution and the  existence of a finite-temperature phase transition relies entirely on numerical simulations \cite{palassini1999universal}. 

Even though criticality in a physical system can emerge only in the thermodynamic limit  \cite{lee1952statistical,yang1952statistical}, in laboratory and numerical experiments the system size is always finite: singularities of physical observables are smeared out and  replaced by  smooth maxima. 
In order to characterize the critical point of finite-size systems, a suitably-defined pseudo-critical temperature must be introduced, e.g. the temperature at which such maxima occur. 
In finite-size systems with quenched disorder such a  pseudo-critical temperature  is a random variable depending on the realization of the disorder.
The characterization of the distribution of the pseudo-critical point and of its scaling properties is  still an open problem which
draw the attention of physicists since the very first works of Harris \cite{harris1974effect,aharony1996absence,wiseman1998finite,bernardet2000disorder,igloi2007finite}. Further  studies of such distributions in spin glasses have been performed  in a recent work \cite{castellana2011role}, where some of the authors showed a connection between the fluctuations of the pseudo-critical temperature of the SK model and the theory of of Extreme Value Statistics (EVS) of correlated random variables. 

The EVS of Independent Identically Distributed (IID) random variables is a well-established problem: a fundamental result \cite{galambos1978asymptotic} states that the limiting Probability Distribution Function (PDF) of the maximum of IID random variables belongs to three families of distributions: the Gumbel, Fr\'echet or Weibull distribution.  
Much less is known about  EVS of correlated random variables. 
A noteworthy case of an EVS distribution of correlated random variables that has been recently discovered is the Tracy-Widom (TW) distribution \cite{tracy2002proceeding}, describing the fluctuations of the largest eigenvalue of a Gaussian random matrix. 
The TW distribution has been found to describe the fluctuations of observables of a broad number of physical and mathematical models, like the longest common sequence in a random permutation \cite{baik1999distribution}, directed polymers in disordered media \cite{johansson2000shape} and polynuclear growth models \cite{prahofer2000universal}, which can be described by the Kardar-Parisi-Zhang equation \cite{kardar1986dynamic,calabrese2011exact}. Recently the TW distribution has been found to describe the conductance fluctuations in two- and three- dimensional Anderson insulators \cite{somoza2007universal,monthus2009statistics} and
measured in  growing interfaces of liquid-crystal turbulence \cite{takeuchi2010universal, takeuchi2011growing} experiments.

In this Letter we study the distribution of the pseudo-critical temperature in the SK and in the EA model by means of numerical simulations. Our numerical findings show that the fluctuations of the pseudo-cirtical temperature of the SK model both with Gaussian and binary couplings are described by the TW distribution. This result suggests that the features of the fluctuations of the pseudo-critical temperature are universal, i. e.  stable with respect to the distribution of the disorder. To our knowledge, this is the first time that the ubiquitous TW distribution is shown to play a role in  spin glasses. Moreover, our numerical analysis shows that the fluctuations of the pseudo-critical point of the EA model are described by the Gumbel distribution. These two results shade light on the role played by EVS in spin glasses.

To pose the problem, let us consider a system of $N$ spins $S_i = \pm 1$ located at the vertices   of a graph, interacting via the Hamiltonian
$
H[ \vec S ] = - \sum_{(i,j)}J_{ij}S_i S_j  ,
$
where the sum runs over the interacting spin pairs $(i,j)$. For the SK model with Gaussian couplings (GSK) and for the SK model with binary couplings (BSK) the interacting spin pairs are all the distinct pairs. The couplings  $J_{ij}$ are IID Gaussian random variables with zero mean and variance $1/N$ for the GSK model \cite{sherrington1975solvable}, and are equal to $\pm1/\sqrt N$ with equal probability for the BSK model \cite{coluzzi2000energy}. For the EA model the interacting spin pairs are the nearest-neighbors pairs on a three-dimensional cubic lattice with periodic boundary conditions, and $J_{ij}$ are IID random variables equal to $\pm 1$ with equal probability \cite{edwards1975theory}. For the BSK and EA model, the binary structure of the couplings allowed for the use of an efficient asynchronous multispin-coding simulation technique \cite{palassini1999universal}, yielding an extensive number of disorder samples and system sizes. 

\begin{figure}
\vspace{5mm}
\includegraphics[scale=0.85]{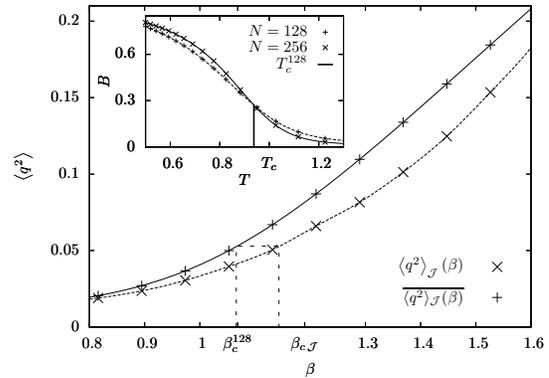}
\caption{
Square value of the overlap $\langle q^2 \rangle_{\mathcal J }$  for a sample $\mathcal J$ (dashed curve) for the BSK model with $N = 128$, its average $\overline{\langle q^2 \rangle_{\mathcal J }}$ over  the samples $\mathcal J$ (solid curve) as a function of the inverse-temperature $\beta$, and critical temperatures $\beta_{c }^ N$ and $\beta_{c \, \mathcal J}$.  The dashed vertical lines depict the definition  (\ref{beta_c_j}) of $\beta_{c \, \mathcal J}$. Inset: Binder parameter $B$ as a function of the temperature $T$ for $N=128,256$ and average pseudo-critical temperature $T_{c }^{ 128}$,  with $T_c^N \equiv 1 / \beta_c^N$.}
\label{plot_def}
\end{figure}

Let us now define the physical observables used to carry on the numerical analysis of the problem. Given two real  spins replicas  $\vec {S}^1, \vec{S}^2$, their mutual overlap $q \equiv \frac{1}{N} \sum_{i=1}^N S^1_i S^2_i $ is a physical quantity characterizing the spin-glass transition in the thermodynamic limit \cite{MPV,palassini1999universal}: $\overline{\langle q^2 \rangle_{\mathcal J}(\beta)} = 0 \text{ if } \beta<\beta_c,  \, \overline{\langle q^2 \rangle_{\mathcal J}(\beta)} > 0 \text{ if } \beta>\beta_c$,
where $\left\langle \cdots \right\rangle_{\mathcal J}$ denotes the thermal average performed with the Boltzmann weight defined by the Hamiltonian $H[\vec S]$,  $\beta \equiv 1/T$ is the inverse temperature, and $\overline{\cdots }$ stands for the average over quenched disorder $\mathcal{J} \equiv \{ J_{ij} \}_{ij}$.
The finite-size inverse pseudo-critical temperature $\beta_{c \, \mathcal{J}}$ of a sample with a realization $\mathcal{J}$  of the disorder can be defined as the value of $\beta$ at which $\langle q^2 \rangle_{\mathcal J}(\beta)$ \textit{significantly differs from zero}, i. e. becomes critical.
This qualitative definition is made quantitative by setting 
\be\label{beta_c_j}
\langle q^2 \rangle_{\mathcal J}(\beta_{c \,  \mathcal J}) = \overline{\langle q^2 \rangle_{\mathcal J}(\beta_{c}^{ N})}. 
\ee
Both for the GSK and BSK model, $\beta_{c}^N$ is chosen to be the average critical temperature at size $N$, which is defined  
 as the temperature at which the Binder ratio  $B \equiv 1/2 ( 3-\overline {\langle q^4\rangle_{\mathcal J}}/\overline {\langle q^2\rangle_{\mathcal J}}^2)$  of a system of size $N$ equals the Binder ratio of a system of size $2N$. For the EA model we simply take $\beta_{c  }^N$ to be equal to the infinite-size critical temperature $\beta_c = 0.855$ \cite{nakamura2003weak}, because in this case the Binder ratios cross at a temperature which is very close to the infinite-size critical temperature $\beta_c$. The definition (\ref{beta_c_j}) and ${\beta_c^N}$ are qualitatively depicted in Fig. \ref{plot_def}. The distribution of $\beta_{c \, \mathcal{J}}$ can be characterized by its mean $\overline{\beta_{c \, \mathcal{J}}}$, its variance $\sigma^2_{\beta \, N} \equiv \overline{  \beta_{c \, \mathcal{J}}^2} - \overline{\beta_{c\,  \mathcal{J}}}^2   $ and by the PDF $p_N(x_{\mathcal{J}})$ of the natural scaling variable $ x_{\mathcal{J}} \equiv ( \beta_{c \,  \mathcal{J}} - \overline{ \beta_{c  \, \mathcal{J}} } )/ \sigma_{\beta \, N}$.
We can expect that, to leading order in $N$, $\sigma_{\beta \, N} \sim N^{-\phi} $ and that for large $N$ $p_N(x_{\mathcal J})$  converges to a nontrivial limiting PDF $p_{\infty}(x_{\mathcal{J}})$.

\begin{figure}
\vspace{5mm}
\includegraphics[scale = .85]{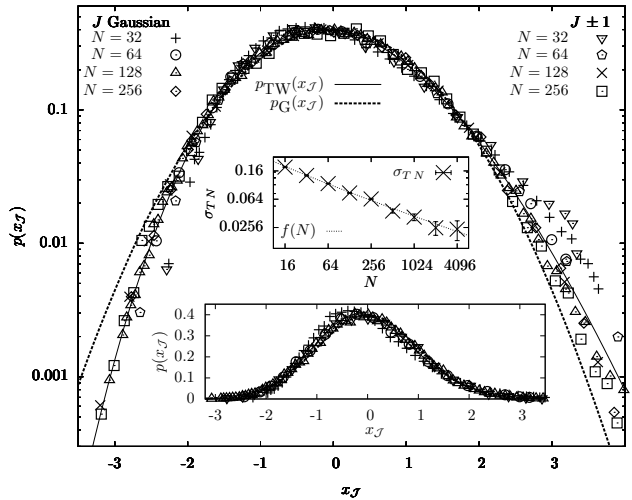}
\caption{
 Distribution of the pseudo-critical point both for the SK model with Gaussian couplings (GSK) and for the SK model with binary couplings (BSK). PDF $p_N (x_{\mathcal J})$  of the rescaled critical temperature $x_\mathcal J$ for systems sizes $N=32,64,128,256$ with  $1.6 \times 10^4 \leq S \leq 4.7 \times 10^4$ (GSK) and $2.9 \times 10^4 \leq S \leq 9.8 \times 10^4$ (BSK) disorder samples,  Tracy-Widom distribution $p_{\textrm{TW}}(x_{\mathcal J})$ (solid curve) and Gaussian distribution $p_{\textrm{G}}(x_{\mathcal J})$ (dashed curve), both with zero mean and unit variance. The plot has no adjustable parameters, and is in logarithmic scale to highlight the behavior of the distributions on the tails.
Top inset: width $\sigma_{T \, N}$ for the BSK as a function of $N$ and fitting function $f(N) = a N^{-\phi}  + b N^{-2\phi}$, yielding $\phi =  0.34        \pm 0.05 $. 
Bottom inset: same plot as in the main plot in linear scale. 
}
\label{plot_sk}
\end{figure}

\textit{Sherrington-Kirkpatrick Model} - Let us start discussing the distribution of $\beta_{c \, \mathcal J}$ for the GSK and BSK model. Monte Carlo (MC) simulations have been performed  with Parallel Tempering (PT) for system sizes $N = 32,64,128,256$ (GSK) and $N = 16, 32, 64, 128,256,512, 1024, 2048,4096$ (BSK),   allowing for a numerical computation of $\langle q^2  \rangle_{\mathcal J}$ and so of $\beta_{c \, \mathcal J}$ for several samples $\mathcal J$. 
The data show that as the system size $N$ is increased, $\overline{ \beta_{c \, \mathcal J} }$ approaches $\beta_c$. 
Setting $T_{c \, \mathcal J} \equiv 1/\beta_{c \, \mathcal J}$, $\sigma^2_{T \, N} \equiv \overline{  T_{c \, \mathcal{J}}^2} - \overline{T_{c\,  \mathcal{J}}}^2 \sim N^{-\phi}   $, the power law fit of $\sigma_{T \, N}$ shown in Fig. \ref{plot_sk} gives the value of the scaling exponent $\phi = 0.31    \pm 0.07$ (GSK) and $\phi = 0.34        \pm 0.05$ (BSK). These values of $\phi$ are both consistent with the value $\phi = 1/3$ one would expect from scaling arguments by considering the variable $y \equiv N^{1/3}(T-T_c)$ \cite{ParisiRitort93a}. 

The PDF $p_N$ of the rescaled variable $x_{\mathcal{J}}$ is depicted in Fig. \ref{plot_sk}. The curves $p_N(x_{\mathcal J})$ collapse quite satisfyingly  indicating that we are close to the asymptotic regime $N \rightarrow \infty$. Even though one could naively expect the fluctuations of the pseudo-critical point to be Gaussian,  Fig. \ref{plot_sk} shows that this is not the case.

 To understand this fact, let us recall the analysis proposed in a recent work \cite{castellana2011role} by some of the authors. In order to study the sample-to-sample fluctuations of the pseudo-critical temperature one uses the Thouless-Anderson-Palmer approach for the SK model. In the TAP approach a free energy  function of the local magnetization is built up for any sample  $\mathcal J$ of the disorder, and its  Hessian matrix $H_{ij}$ calculated at the paramagnetic minimum is a random matrix in the GOE ensemble. In the thermodynamic limit, the spectrum of $H_{ij}$ is described by the Wigner semicircle, centered in $1+ \beta^2$ and with radius $2 \beta$. The critical temperature $\beta_c = 1$ of the SK model is identified as the value of $\beta$ such that the minimal eigenvalue of $H_{ij}$ vanishes. In \cite{castellana2011role} the fluctuations of the pseudo-critical temperature are investigated in terms of the fluctuations of the minimal eigenvalue of $H_{ij}$. One  introduces a definition of pseudo-critical temperature $\hat{\beta}_{c \, \mathcal J}$, which is different from that considered in the present work. The finite-size fluctuations of $\hat{\beta}_{c\, \mathcal J}$  are found to be described by  the relation
$
\hat{\beta}_{c \, \mathcal{J}} = \beta_c -  \chi_{\mathcal J}/(2N^{2/3}),
$
where $\chi_{\mathcal J}$ is distributed according to the TW distribution in the high-temperature region  $\hat{\beta}_{c \, \mathcal J} < 1$. According to Fig. \ref{plot_sk}, MC simulations confirm this analysis: the limiting distribution of $p_N(x_{\mathcal J})$ is described with good accuracy by the TW distribution in the high-temperature regime $\beta_{c \, \mathcal J} < 1$ ($x_{\mathcal J}<0$). The TW distribution is robust with respect to the choice of the disorder distribution and to the definition of pseudo-critical temperature.  On the other hand, since the exponent $\phi$ obtained from MC simulations is not compatible the exponent $2/3$ of $\hat{\beta}_{c \, \mathcal J}$, we conclude that the scaling exponent is definition-dependent \cite{castellana2011role,billoire2011finite}.

\begin{figure}
\vspace{5mm}
\includegraphics[width=8.5cm]{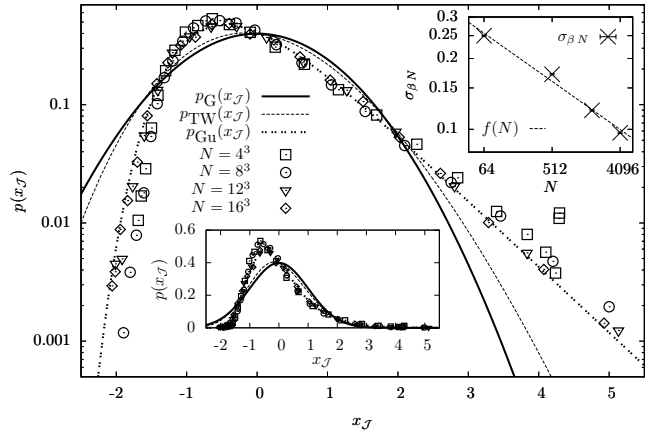}
\caption{
Distribution of the pseudo-critical point for the EA model. PDF $p_N (x_{\mathcal J})$  of the rescaled critical temperature $x_{\mathcal J}$ for systems  sizes $N=4^3,8^3,12^3,16^3$ with  $2.4 \times 10^4 \leq S \leq 3.2 \times 10^4 $ disorder samples,  Gaussian distribution $p_{\textrm{G}}(x_{\mathcal J})$ (solid curve), Tracy-Widom distribution $p_{\textrm{TW}}(x_{\mathcal J})$ (dashed curve) and Gumbel distribution $p_{\textrm{Gu}}(x_{\mathcal J})$ (dotted  curve), all with zero mean and unit variance.
The plot has no adjustable parameters, and is in logarithmic scale to highlight the behavior of the distributions on the tails. Top inset: width $\sigma_{\beta \, N}$ as a function of $N$, and fitting function $f(N) = a \, N^{-\phi}$, with scaling exponent $\phi =  0.23     \pm 0.03$ . Bottom inset: same plot as in the main plot in linear scale. 
}
\label{plot_ea}
\end{figure}

\textit{Edwards-Anderson Model} - The same analysis has been performed for the three-dimensional EA model. Physical observables have been computed with PT  for system sizes $N = L^3$ with $L = 4,8,12,16$. Similarly to the SK model, the width  $\sigma_{\beta \, N}$ of the distribution of the pseudo-critical point $\beta_{c \, \mathcal J}$ shrinks to zero as the system size $N$ is increased: a power law fit $\sigma_{\beta \, N} = a \, N^{-\phi}$ gives the value of the scaling exponent  $\phi =  0.23         \pm 0.03 $ (inset of Fig. \ref{plot_ea}). 
The PDFs $p_N(x_{\mathcal J})$ of the rescaled critical temperature seem to have a finite limit as $N$ is increased, as depicted in Fig. \ref{plot_ea}, and this limit coincides with the Gumbel distribution.
Both $\phi$ and  the PDF have the following interesting features.  
As far as the exponent $\phi$ is concerned, we recall \cite{aharony1996absence}  that  for systems known to be governed by a random fixed point like the EA model it was predicted that the scaling exponent satisfies $1/\phi = d \nu$, where $d$ is the dimensionality of the system. The value of the critical exponent $\nu = 1.8 \pm 0.2$ for the EA model is known from numerical simulations \cite{palassini1999universal},  yielding a value of $\phi = 0.19     \pm 0.02 $ which is  compatible with that  measured  from the fluctuations of the critical temperature.
As far as the limiting distribution $p_\infty(x_{\mathcal J})$ is concerned, we recall that  \cite{vojta2006rare}  a disordered system like the EA behaves as an ensemble of independent sub-systems $\mathcal{S}_1, \ldots, \mathcal{S}_M$, where each sub-system $\mathcal{S}_i$ has a random local critical temperature $\beta_c^i$, the local critical temperatures $\{ \beta_c^i \}_i$ being  IID random variables depending on the \emph{local} realization of the disorder.  We can argue that, for a single realization of the disorder $\mathcal J$,  the pseudo-critical temperature $\beta_{c \, \mathcal J}$  results from the fact that $\beta$  has to be taken large  enough to bring all of the sub-systems $\{ \mathcal{S}_i\}_i$ to criticality. Thus, $\beta_{c \, \mathcal J}$ is the maximum over the ensemble of the local critical temperatures $ \beta_{c \, \mathcal J} = \max_i\beta_c^i$. If this picture is correct, $\beta_{c \, \mathcal J }$ is distributed according to one of the the EVS limiting distributions of independent variables \cite{galambos1978asymptotic}: the Gumbel, Fr\'echet, or Weibull  distribution. Assuming that the distribution of $\beta_c^i$ decays exponentially for large $\beta_c^i$, the distribution of $\beta_{c \, \mathcal J}$ is the Gumbel one. We want to stress that this argument would not hold for the SK model, where there is no geometric structure.
 
\textit{Conclusions} - In this Letter we have performed a numerical analysis of the distribution of the pseudo-crtical temperature in two mean-field spin glasses, the Sherrington-Kirkpatrick  model with Gaussian couplings (GSK) and with binary couplings (BSK), and in a short-range spin glass, the Edwards-Anderson (EA) model. The analysis for the GSK and BSK model shows that the distribution of the  pseudo-critical temperature in the high temperature phase is described with good accuracy by the Tracy-Widom (TW) distribution, as suggested by an analytical prediction previously published by some of the authors \cite{castellana2011role}. To our knowledge, this is the first time that the TW distribution is shown to play a role in spin glasses. The fact that both the GSK and BSK yield the TW distribution suggests that the TW distribution is universal with respect to the bonds' distribution. 

The analysis pursued for the three-dimensional EA model shows that the liming distribution of the pseudo-critical temperature  is the Gumbel distribution. An argument to understand this result has been proposed. These two numerical analyses put in evidence a  connection between the critical regime of spin-glass models and the Extreme Value Statistics theory which has never been proposed heretofore.

The present Letter opens several perspectives. 
As far as the SK model is concerned, we recall that the TW distribution describes typical fluctuations of the maximal eigenvalue of a Gaussian Orthogonal Ensemble random matrix, while the large deviations regime of these fluctuations has been studied only recently \cite{dean2006large}. It would be interesting to study numerically the large deviations regime of the fluctuations of the critical temperature, where the distribution of the pseudo-critical point could be described by the large deviations function derived in \cite{dean2006large}. 
It would be also interesting to consider the case where the couplings $J_{ij}$ are Gaussian with a positive bias $J_0$ \cite{NishimoriBook01}.  Depending on the value of $J_0$,  the SK model has a   phase transition from a paramagnetic to a spin-glass phase or from a  ferromagnetic to a mixed phase \cite{NishimoriBook01}: it would be interesting to investigate both analytically and numerically the fluctuations of these pseudo-critical points.  
Moreover, in order to bridge the gap between a mean-field and a short-range interactions regime, it could be interesting to investigate the fluctuations of the pseudo-critical temeperature in spin-glass models with tunable long-range interactions, like that introduced in \cite{kotliar1983one}. 
As far as the EA model is concerned, it would be interesting to test experimentally the scenario found here in  $\text{Fe}_{0.5}\text{Mn}_{0.5}\text{TiO}_3$ \cite{gunnarson1991static} or $\text{Eu}_{0.5}\text{Ba}_{0.5}\text{MnO}_3$ \cite{nair2007critical} spin-glass materials. Indeed, ac-susceptibility measurements in these systems show  \cite{gunnarson1991static}  that the spin-glass critical temperature can be identified as the temperature where the susceptibility has a cusp. Accordingly, the pseudo-critical point  could be  identified and measured, and
one could test whether the resulting rescaled pseudo-critical point distribution converges to the Gumbel distribution as the system size is increased.
\paragraph*{Acknowledgments}
We are glad to thank G. Parisi, A. Rosso and P. Vivo for interesting discussions and suggestions. We also acknowledge support from the D. I. computational center of University \textit{Paris Sud} and from the LPTMS cluster.

\bibliography{bibliography}

\end{document}